\documentclass[a4paper,11pt]{article}
\usepackage{pos}
\usepackage{float}
\usepackage{siunitx}
\usepackage{booktabs}
\usepackage{mathtools}
\usepackage{physics}
\usepackage{bbm}
\usepackage[noabbrev, nameinlink]{cleveref}
\creflabelformat{equation}{#2\textup{#1}#3}
\usepackage{csquotes}
\usepackage{sidecap}
\sidecaptionvpos{figure}{m}
\sidecaptionvpos{table}{m}
\usepackage{tikz}
\usetikzlibrary{arrows.meta}

\title{Parallel Tempered Metadynamics}
\ShortTitle{Parallel Tempered Metadynamics}

\author*{Timo Eichhorn}
\author*{Gianluca Fuwa}
\author{Christian Hoelbling}
\author{Lukas Varnhorst}

\affiliation{Department of Physics, University of Wuppertal, Gaußstraße 20, D-42119 Wuppertal, Germany}

\emailAdd{timo.eichhorn@protonmail.com}
\emailAdd{gianluca.fuwa@uni-wuppertal.de}

\abstract{When approaching the continuum limit in lattice QCD or other theories in a setup with topological sectors, conventional update algorithms experience a particularly severe form of critical slowing down that is caused by high action barriers between the sectors. The qualitative scaling behavior of this critical slowing down appears to be universal across different update algorithms and (gauge) actions. We demonstrate that a combination of Metadynamics with parallel tempering, along with other modifications, can significantly reduce autocorrelation times while avoiding the need for reweighting. We also discuss strategies to extend the methods to QCD simulations with dynamical fermions, and present first results for $N_f = 2$ simulations at unphysical pion masses.}

\FullConference{The 41st International Symposium on Lattice Field Theory (LATTICE2024)\\
28 July - 3 August 2024\\
Liverpool, UK\\}

\begin{document}
\maketitle
\newcommand\thefont{\expandafter\string\the\font}

\section{Introduction}
Virtually all current large-scale simulations of lattice QCD are Markov chain Monte Carlo (MCMC) simulations based on the Hybrid Monte Carlo (HMC) algorithm \cite{Duane:1987de} or one of its variants, with fermions represented indirectly using pseudofermions \cite{Weingarten:1980hx}. Under these general assumptions, the cost of sampling independent configurations can be approximately described by a scaling law involving three terms:
\begin{equation}
    \textnormal{cost} \propto V^{z_{V}} m_{\pi}^{-z_\pi} a^{-z_a}.
\end{equation}
This does not take into account the computation of observables, which may exceed the cost of generating configurations but is beyond the scope of this discussion.
\paragraph{Volume scaling}
The volume scaling is caused by the need to decrease the integration step size in the HMC with increasing volumes to achieve constant acceptance rates in the accept-reject step, or equivalently, constant variances of the energy violations during trajectories. The details depend on the integrator, but even with the simplest choice of a (second order) leapfrog integrator, the overall scaling behavior is relatively mild with an exponent $z_V = 1 + 1/4$ \cite{Creutz:1988wv, Gupta:1988js}. More generally, for an integrator of order $n$, the exponent is given by $z_V = 1 + 1/2n$ \cite{Creutz:1989wt}. Ultimately, it is difficult to envisage an algorithm achieving better than linear volume scaling, as the computational cost of updating the degrees of freedom in the system necessarily grows with the system size.

\paragraph{Pion/quark mass scaling}
The scaling in the pion mass or the square root of the quark mass can be attributed to the presence of smaller eigenvalues of the Dirac operator. As a consequence, the condition number of the Dirac operator increases, and the solver requires more iterations until it converges. Additionally, the small eigenvalues of the Dirac operator tend to fluctuate strongly, which leads to increased variations in the fermion forces. This, in turn, means that smaller integrator step sizes are required to avoid instabilities. Historically, simulations at physical pion masses and reasonable lattice spacings and volumes seemed to be infeasible up until the early 2000s \cite{Ukawa:2002pc}, but various algorithmic advancements \cite{Hasenbusch:2001ne, Urbach:2005ji, Luscher:2003vf, Luscher:2003qa, Luscher:2004pav, Sexton:1992nu, DeGrand:1990dk, Clark:2006fx, Luscher:2007se, Luscher:2007es, Brannick:2007ue, Babich:2010qb, Osborn:2010mb, Frommer:2013fsa, Brower:2020xmc} over the years have made those simulations possible. Finally, unlike the other two scaling terms, the pion mass is typically bounded from below around its physical value. Most calculations are targeted around the physical point and do not require chiral extrapolations, which prevents the mass scaling problem from becoming excessively severe.

\paragraph{Lattice spacing scaling}
Finally, the scaling in the lattice spacing is a consequence of the fact that MCMC algorithms are used to sample field configurations. Naively, one might expect a Langevin-like quadratic scaling of the autocorrelation times with the inverse lattice spacing when using the HMC algorithm \cite{Luscher:2011qa}, but the presence of topological sectors seems to lead to a significantly worse scaling behavior and large autocorrelation times for both quenched and dynamical simulations \cite{Alles:1996vn, Orginos:2001xa, DeGrand:2002vu, Noaki:2002ai, Aoki:2005ga, Schaefer:2009xx, Schaefer:2010hu}. Empirically, the scaling of the integrated autocorrelation times has been found to be compatible with a critical exponent $z_a \approx$ \numrange{5}{6}, or alternatively an exponential scaling in the lattice spacing. Although this issue was not a practical concern for a long time, it has persisted since the earliest days of lattice QCD and affects not only topological observables, but also the correctness of non-topological quantities. There has been progress in recent years in addressing both general critical slowing down \cite{DUANE1986143, Duane:1988vr, Davies:1989vh, Cossu:2017eys, Zhao:2018jas, Sheta:2021hsd, Nguyen:2021zgx, Horsley:2023fhj, Ostmeyer:2024amv, Huo:2024lns, Jung:2024nuv, Horsley:2025rof, Fields:2025ydm, Christ:2025pih, Lundstrum:2024aoe, Luscher:2009eq, Engel:2011re, Bacchio:2022vje, Boyle:2022xor, Matsumoto:2023akw, Chamness:2024fob, Yamamoto:2025imx, Endres:2015yca, Detmold:2016rnh, Detmold:2018zgk, Howarth:2023bwk, Pawlowski:2018qxs, Albergo:2019eim, Kanwar:2020xzo, Boyda:2020hsi, DelDebbio:2021qwf, Foreman:2021ljl, Finkenrath:2022ogg, Caselle:2022acb, Albergo:2022qfi, Singha:2022lpi, Abbott:2022zhs, Abbott:2022hkm, Komijani:2023fzy, Albandea:2023wgd, Abbott:2023thq, Singha:2023xxq, Caselle:2023mvh, Wang:2023exq, Albandea:2023ais, Foreman:2023ymy, Finkenrath:2024tdp, Abbott:2024knk, Abbott:2024mix, Jin:2023shj, Caselle:2024ent, Bulgarelli:2024brv, Komijani:2025yjz, Abbott:2025kvi, Zhu:2025pmw} and topological freezing \cite{FUCITO1984230, Smit:1987fq, Dilger:1992yn, Dilger:1994ma, Durr:2012te, Albandea:2021lvl, Luscher:2011kk, Mages:2015scv, Luscher:2017cjh, deForcrand:1997fm, Laio:2015era, Bonati:2017nhe, Eichhorn:2023uge, Bonati:2017woi, Hasenbusch:2017unr, Bonanno:2024zyn, Bonanno:2024udh, Vadacchino:2024lob, Cossu:2021bgn, Foreman:2021rhs}, although the majority of the more widely used methods circumvent the latter problem.
\section{Critical slowing down and topological freezing}
The particularly unfavorable scaling with the lattice spacing is not unique to QCD and has been observed in other topologically non-trivial theories, including two-dimensional $\mathrm{CP^{N-1}}$ models \cite{Campostrini:1992ar, Vicari:1992jy, DelDebbio:2004xh, Flynn:2015uma} and U(N) gauge theories \cite{Kanwar:2020xzo, Albandea:2021lvl, Eichhorn:2021ccz, Rouenhoff:2024geh}, as well as the Schwinger model \cite{Albandea:2021lvl, Albergo:2022qfi, Finkenrath:2022ogg, Finkenrath:2024tdp} and four-dimensional SU(N) gauge theories \cite{DelDebbio:2001sj, DelDebbio:2002xa, Schaefer:2009xx, Schaefer:2010hu, McGlynn:2014bxa, Eichhorn:2023uge}. A common feature of these theories in the continuum is that the field space is divided into disconnected sectors separated by infinitely high action barriers. On the lattice with periodic boundary conditions, these barriers remain finite but grow as the continuum limit is approached. This becomes problematic for conventional update algorithms, which traverse the configuration space in a pseudo-continuous fashion. Relevant regions of the configuration space are separated by low probability density regions, and transitions between topological sectors become increasingly unlikely. For very fine lattice spacings below $\sim \SI{0.04}{\femto\meter}$, a considerable amount of computational effort is required to correctly sample different topological sectors, and the ergodicity of the simulation is threatened. For a more detailed discussion of the problem and various solutions that have been proposed, we refer to recent review articles \cite{Finkenrath:2023sjg, Kanwar:2024ujc, Boyle:2024nlh, Finkenrath:2024ptc}.

\begin{figure}[H]
    \centering
    \includegraphics[width=0.49\linewidth]{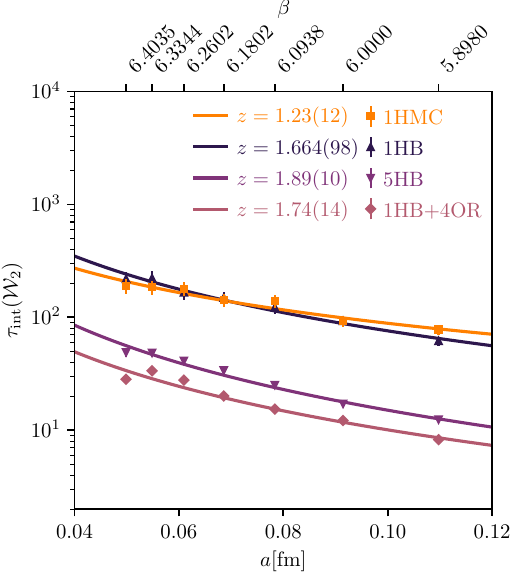}
    \includegraphics[width=0.49\linewidth]{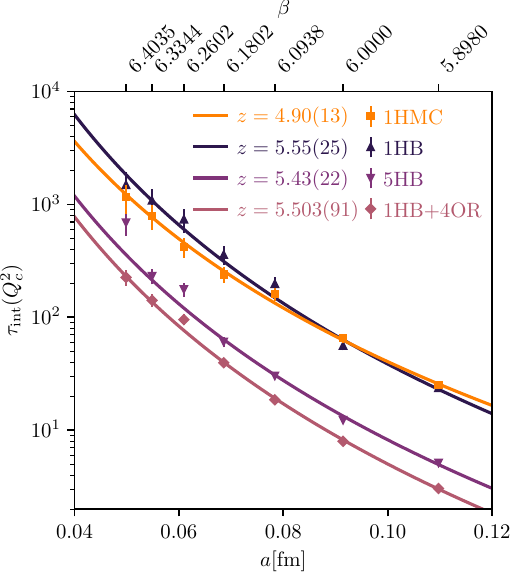}
    \caption{Scaling of integrated autocorrelation times with the lattice spacing for different update algorithms in four-dimensional SU(3) gauge theory, using the Wilson gauge action (figure taken from \cite{Eichhorn:2023uge}). The data are compatible with power-law fits in the lattice spacing.}
    \label{fig:tau_int_scaling_algorithms}
\end{figure}

\Cref{fig:tau_int_scaling_algorithms} shows the scaling of the integrated autocorrelation times for $2 \times 2$ Wilson loops and the squared clover-based topological charge for unit length trajectory HMC updates, heat bath updates \cite{PhysRevLett.43.553, Creutz:1980zw, Fabricius:1984wp, Kennedy:1985nu, Cabibbo:1982zn}, and a combination of heat bath and overrelaxation updates \cite{Adler:1981sn, Creutz:1987xi, Brown:1987rra}. Both observables were measured after 31 stout smearing steps with a smearing parameter $\rho = 0.12$.
\noindent
For the $2 \times 2$ Wilson loops, the increase of the integrated autocorrelation times is compatible with a critical exponent $z_a \approx $ \numrange{1}{2}, which is consistent with the behavior observed for larger $4 \times 4$ and $8 \times 8$ loops (not shown here). In contrast, the scaling of the integrated autocorrelation times of the squared topological charge is compatible with a critical exponent $z_a \approx 5$, and their magnitudes start to surpass those of other observables for lattice spacings below $a \approx \SI{0.08}{\femto\meter}$. Among the update algorithms considered here, the combination of heat bath and overrelaxation updates appears to be the most efficient in terms of both update sweeps and computational cost. However, while the values may vary by approximately an order of magnitude across the different update schemes, the overall scaling behavior remains consistent. A more detailed discussion of the results can be found in our previous work \cite{Eichhorn:2023uge}.

\begin{figure}[H]
    \centering
    \includegraphics[width=0.49\linewidth]{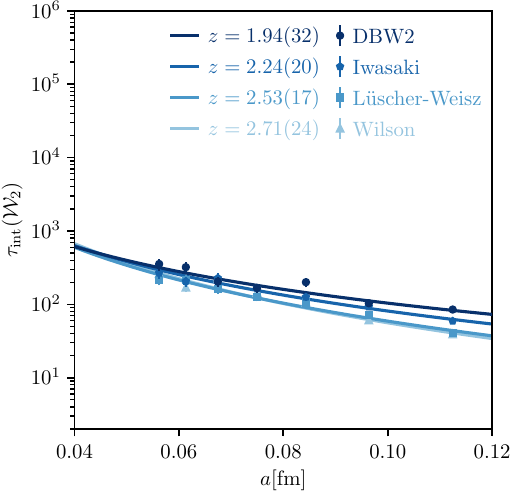}
    \includegraphics[width=0.49\linewidth]{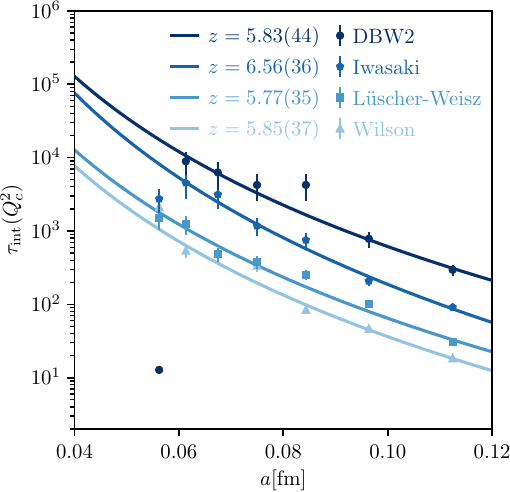}
    \caption{Scaling of integrated autocorrelation times with the lattice spacing for different gauge actions using the HMC in four-dimensional SU(3) gauge theory \cite{Eichhorn_Thesis}. The DBW2 and Iwasaki results for the finest lattice spacings were excluded from the fits, as there were not sufficient samples for an accurate estimate. Qualitatively similar scaling behavior was also observed for local update algorithms (1HB+4OR), albeit with smaller autocorrelation times.}
    \label{fig:tau_int_scaling_actions}
\end{figure}

Similarly, \Cref{fig:tau_int_scaling_actions} shows the scaling of the integrated autocorrelation times for the corresponding observables after 30 smearing steps, using the HMC algorithm, for different gauge actions within a one-parameter family that includes plaquettes and planar $1 \times 2$ Wilson loops.
\noindent
While the autocorrelation times of the $2 \times 2$ Wilson loops are compatible with each other for the different gauge actions, there is more than an order of magnitude of difference for the squared topological charge between the action with the shortest and the longest observed autocorrelation times. Among the gauge actions considered here, the problem is most severe for the DBW2 action \cite{Takaishi:1996xj}, followed by the Iwasaki action \cite{Iwasaki:1983iya}, the Lüscher-Weisz action \cite{Luscher:1984xn}, and finally the Wilson gauge action \cite{Wilson:1974sk}. As one would expect for the DBW2 and the Iwasaki actions \cite{Orginos:2001xa, DeGrand:2002vu, Noaki:2002ai, Aoki:2005ga, Schaefer:2010hu}, the problem is severe enough that there are not sufficient samples to accurately estimate the integrated autocorrelation times of $Q_c^2$ at the finest lattice spacing.

It is important to emphasize that the results shown in \Cref{fig:tau_int_scaling_algorithms} and \Cref{fig:tau_int_scaling_actions} do not imply that Wilson loops are entirely unaffected --- their coupling to the slow topological modes may simply be too weak to detect at the given level of precision. Finally, the uncertainties on the dynamical critical exponents are presumably underestimated, and should be interpreted as a qualitative guide rather than a precise determination of the continuum scaling behavior.
\section{Parallel Tempered Metadynamics}
Metadynamics \cite{Laio_2002, Laio:2015era} is an enhanced-sampling method that introduces an adaptive bias potential, $V_t(s(U))$, which evolves over simulation time $t$ and depends on a set of collective variables (CVs) $\{s_i\}$. In principle, these CVs may be arbitrary functions of the underlying degrees of freedom $\{U\}$ of the system, but when used in combination with the HMC or other molecular-dynamics-based algorithms, they must also be differentiable. During the simulation, the potential is iteratively updated by adding kernel functions $g(s)$, typically Gaussians with height $w_i$ and width $\delta s_i$, centered at the current point in CV space. Over time, the bias potential converges towards the negative free energy as a function of the CVs, up to a constant offset \cite{PhysRevLett.96.090601, PhysRevE.81.055701}. Equivalently, this can be interpreted as flattening the marginal distribution over the CVs. Assuming a setup with $N$ collective variables, the bias potential at simulation time $t$ is given by
\begin{equation}
    V_t(s) = \sum\limits_{t' \leq t} \prod\limits_{i=1}^{N} g\bigl(s_i - s_i(t')\bigr).
\end{equation}
Unbiased expectation values with respect to the original probability distribution can be recovered through reweighting, although correctly handling the time-dependent probability distribution requires some care. Various approaches have been suggested in the literature on how to address this problem \cite{doi:10.1021/acs.jctc.9b00867}, but a simpler alternative is to use a static bias potential derived from one or more prior simulations with dynamic potentials. In this case, the simulation resembles a multicanonical simulation \cite{PhysRevLett.68.9, Bonati:2018blm}.

\begin{figure}[H]
    \centering
    \includegraphics[width=0.85\linewidth]{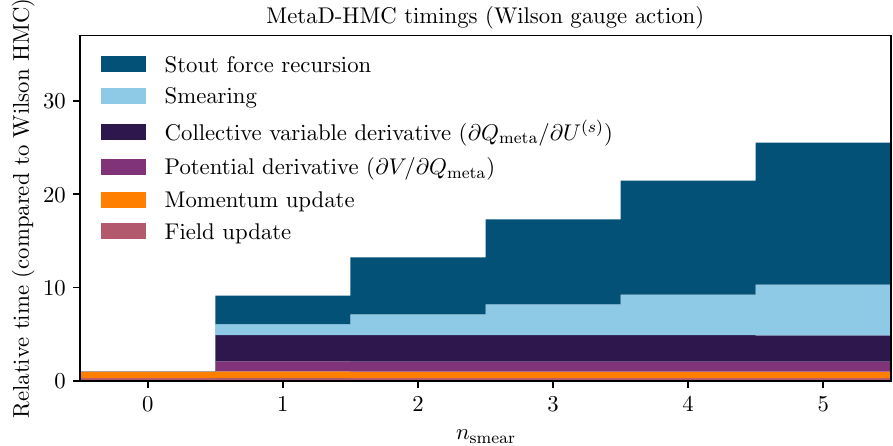}
    \includegraphics[width=0.85\linewidth]{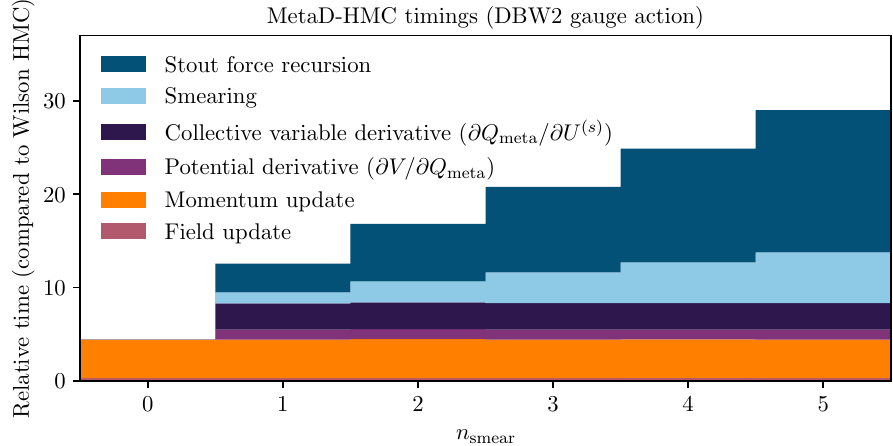}
    \caption{Breakdown of the computational costs for the MetaD-HMC algorithm for four-dimensional SU(3) gauge theory.}
    \label{fig:MetaD_timings}
\end{figure}

For pure gauge theory, we use the clover-based definition of the topological charge as the CV after applying 4 stout smearing steps with a smearing parameter $\rho = 0.12$ to the gauge fields. To distinguish this choice from other definitions of the topological charge, the CV is denoted by $Q_{\textrm{meta}}$. Further details and the rationale behind the choice of CV can be found in \cite{Eichhorn:2023uge}. The bias potential introduces a new force term in addition to the already existing force terms when using the HMC. \footnote{In principle, local update algorithms could be used, but they would be extremely inefficient due to the non-local nature of the smeared CV.} Since we only use a single CV, the new contribution is given by
\begin{equation}
        F_{\mu, \mathrm{meta}}(n) = -\frac{\partial V}{\partial Q_{\mathrm{meta}}} \frac{\partial Q_{\mathrm{meta}}}{\partial U^{(s)}_{\mu_s}(n_s)} \frac{\partial U^{(s)}_{\mu_s}(n_s)}{\partial U^{(s-1)}_{\mu_{s-1}}(n_{s-1})} \dots \frac{\partial U^{(1)}_{\mu_1}(n_1)}{\partial U_{\mu}(n)},
        \label{eq:MetaD_force}
\end{equation}
with implicit summations over the lattice sites $n_i$ and the Lorentz indices $\mu_i$. The derivative of the bias potential with respect to $Q_{\mathrm{meta}}$ is trivial to compute, but the remaining terms are more complicated. In particular, computing the derivative of the smeared gauge fields $U^{(s)}$ after $s$ stout smearing steps with respect to the unsmeared fields $U$ is a nontrivial task, but the procedure corresponds to the force recursion used for stout smeared (fermion) actions \cite{Morningstar:2003gk}. It is computationally advantageous to evaluate the force from left to right, i.e., to start with the derivative with respect to the maximally smeared fields. Still, the computational overhead is considerable for pure gauge theory, as seen in \Cref{fig:MetaD_timings}. In four-dimensional SU(3) gauge theory, our implementation of the Metadynamics-HMC (MetaD-HMC) algorithm with four stout smearing steps is approximately 20 times slower than the standard HMC algorithm when using the Wilson gauge action. For the DBW2 action, the relative overhead is lower --- around a factor of 6 --- due to the increased baseline cost of the momentum updates from the additional $1 \times 2$ Wilson loops. As one might expect, the majority of the overhead stems from the stout force recursion and the smearing. However, in simulations with dynamical fermions, the relative computational overhead of the MetaD-HMC is expected to be significantly smaller --- around \SI{20}{\percent} for the setup discussed in \Cref{sec:4}.
\noindent
In addition to the computational overhead, the reweighting required to obtain unbiased estimators of observables must also be considered, as it can lead to large statistical uncertainties, depending on the overlap between the original and modified probability distributions. In this context, while Metadynamics reduces the autocorrelation times of topological observables \cite{Laio:2015era, Bonati:2017nhe, Eichhorn:2022wxn, Rouenhoff:2022seh}, its benefits may be partially offset by the increased statistical uncertainties introduced by reweighting \cite{Rouenhoff:2022seh, Eichhorn:2023uge}. One contributing factor is the oversampling of configurations with large $\abs{Q}$, which can be avoided by an appropriate modification of the bias potential. As it is expected that the topological charge distribution is well approximated by a Gaussian distribution for sufficiently large volumes, a possible approach is to preserve the relative weights of individual sectors while flattening only the barriers separating them. This can be achieved via singular spectrum analysis \cite{VAUTARD1989395} or similar techniques that separate different contributions to the bias potential.

\begin{figure}[H]
    \centering
    \vspace{-0.4cm}
    \input{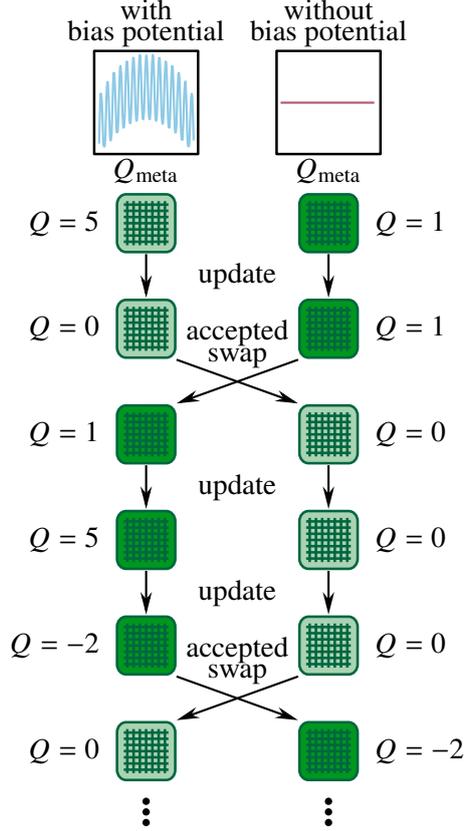}
    \caption{Illustration of the Parallel Tempered Metadynamics (PT-MetaD) algorithm \cite{Eichhorn:2023uge}, where the two columns represent separate simulation streams. The bias potentials are depicted at the top, while subsequent rows represent the time evolution, with different shades distinguishing configurations for better visibility. $Q$ values are illustrative and do not correspond to results from an actual simulation.}
    \label{fig:PT-MetaD_illustration}
\end{figure}

The remaining reweighting overhead comes from the oversampling of configurations in the barrier regions between topological sectors. Since these states are necessary for tunneling between different sectors, they cannot be entirely avoided. However, the reweighting can be circumvented by introducing a second, unbiased simulation stream and periodically proposing swaps between the two streams \cite{Eichhorn:2023uge}, as illustrated in \Cref{fig:PT-MetaD_illustration}. By construction, all measurements performed on the unbiased stream require no reweighting. In this sense, our approach shares similarities with parallel tempering in boundary conditions \cite{Hasenbusch:2017unr}.
\noindent
As long as both streams use the same physical action, the acceptance probability of swaps is determined solely by the difference in the bias potential, since the action difference is given by
\begin{equation}
    \begin{aligned}
        \Delta S^M_t &= \bigl[S^M_{t}(U_1) + S(U_2)\bigr] - \bigl[S^M_{t}(U_2) + S(U_1)\bigr]
        \\           &= V_t(Q_{\mathrm{meta}, 1}) - V_t(Q_{\mathrm{meta}, 2}).
    \end{aligned}
\end{equation}
The numeric indices of the quantities indicate the stream number, $S^M_t$ is the action including the bias potential, and $V_t$ represents the bias potential. In particular, this means that the swap accept-reject steps require no additional inversions of the Dirac operator in simulations with dynamical fermions.
\section{Extension to full QCD}
\label{sec:4}
In some ways, the application of (Parallel Tempered) Metadynamics is inherently better suited for simulations with dynamical fermions than for pure gauge theory. Essentially all large-scale simulations already rely on HMC-based update algorithms, and smeared fermion actions are commonly used, so the stout force recursion required for computing the bias force described in \Cref{eq:MetaD_force} is already implemented in many simulation codes. However, dynamical fermions significantly increase the computational demands, which limits the number of configurations that can be generated. In pure gauge theory, the buildup of the bias potential seems to require $\order{\num{e4}}$ updates, which is no longer feasible in this setting. Therefore, improvements in the efficiency of the bias potential buildup are necessary before the algorithm can be considered viable for large-scale simulations.

While it is possible to adjust the height $w$ and width $\delta Q$ of the Gaussians used to construct the bias potential, this comes with a trade-off between the buildup speed and the accuracy of the potential. Larger values of $w$ and $\delta Q$ can accelerate the buildup in some sense, but lead to larger fluctuations \cite{PhysRevLett.96.090601} and a coarser resolution in CV space, respectively. Instead, we focus on strategies that accelerate the buildup process with little to no sacrifice of smoothness and convergence.

An obvious improvement is to incorporate the parity symmetry present in both pure gauge theory and full QCD into the buildup process \cite{Leinweber:2003sj}. By simultaneously updating the potential at both $Q_{\mathrm{meta}}$ and $-Q_{\mathrm{meta}}$, the buildup is effectively accelerated by a factor of 2.

Another strategy we employ is an extension of standard Metadynamics known as well-tempered Metadynamics \cite{PhysRevLett.100.020603}, where the bias deposition rate decays as the potential grows. In this approach, the update procedure of the potential is modified to
\begin{equation}
    V_{t + 1}(Q) = V_{t}(Q) + \exp(- \frac{V_{t}(Q)}{\Delta T}) w \exp(-\frac{\bigl(Q(t) - Q \bigr)^2}{2 \delta Q^2}) \, ,
\end{equation}
where $\Delta T$ is a newly introduced tunable parameter. In the limit $\Delta T \rightarrow 0$, the method reduces to regular importance sampling without a bias potential, while for $\Delta T \rightarrow \infty$, it becomes equivalent to standard Metadynamics. The damping introduced by $\Delta T$ allows for a larger initial Gaussian height $w$ while still producing a smooth potential after thermalization. However, this modification also causes the potential to converge towards $-\frac{\Delta T}{1 + \Delta T} S(Q)$, up to a constant offset. Therefore, some care must be taken when choosing $\Delta T$, as values that are too small can result in a bias potential that is too weak to overcome the action barriers.

Finally, the most significant speedup in our setup is achieved by using multiple parallel simulation streams (usually referred to as multiple walkers) that share and update the same bias potential \cite{doi:10.1021/jp054359r}. This leads to a near-linear speedup, approximately equal to the number of walkers. The method is also advantageous from a computational perspective, as it can be almost trivially parallelized, requiring only minimal communication between the simulation streams.

For an initial test study, we performed $N_f = 2$ simulations with the DBW2 gauge action and \texttt{4stout} staggered fermions \cite{Kogut:1974ag} at $am = 0.02$ with a smearing parameter of $\rho = 0.125$. The configurations were generated using the Rational Hybrid Monte Carlo (RHMC) algorithm \cite{Kennedy:1998cu, Clark:2003na, Clark:2004cp, Clark:2006wq} on a $V = 16^4$ lattice at $\beta = 1.05$. These parameters were not chosen because of their physical relevance, but rather due to constraints on the available computational resources. An attempt at rough scale setting using the Wilson flow scales $\sqrt{t_0}$ \cite{Luscher:2010iy} and $w_0$ \cite{BMW:2012hcm} suggests that these parameters correspond to an inverse lattice spacing of around \SIrange{3.5}{4.0}{\giga\eV}, or $\sim \SI{0.05}{\femto\meter}$, which implies a volume of approximately $(\SI{0.8}{\femto\meter})^4$. As the collective variable, we use the clover-based topological charge with six steps of stout smearing with a smearing parameter of $\rho = 0.12$, which is a slight increase compared to the pure gauge case.

\begin{figure}[H]
    \centering
    \includegraphics{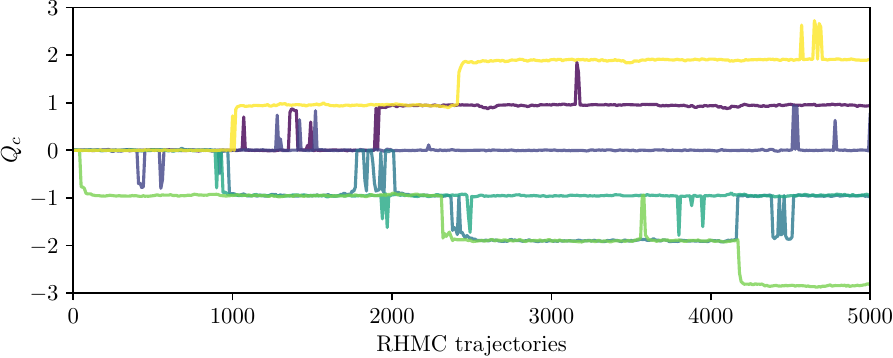}
    \caption{Time series of the clover-based topological charge during the bias potential buildup phase of a Metadynamics run with six walkers in a dynamical QCD simulation.}
    \label{fig:timeseries_multiple_walkers}
\end{figure}

\begin{figure}[H]
    \centering
    \includegraphics{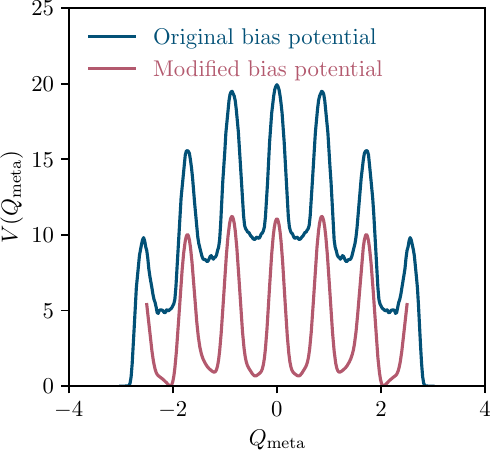}
    \caption{QCD bias potential constructed from six parallel walkers with 5000 RHMC trajectories for each walker. The modified potential was obtained from the original potential using singular spectrum analysis.}
    \label{fig:MetaD_bias_qcd}
\end{figure}

\begin{figure}[H]
    \centering
    \includegraphics{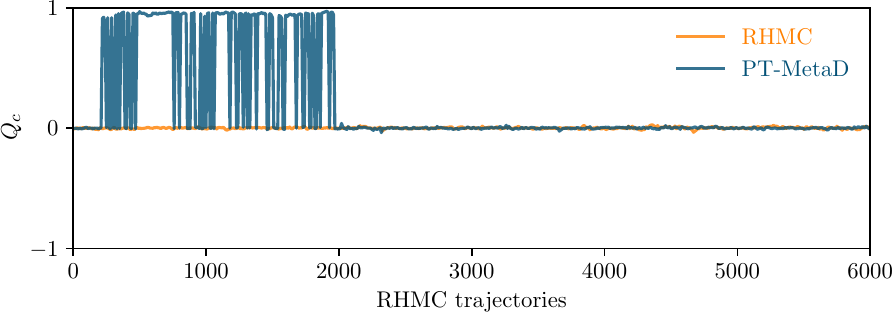}
    \caption{Time series of the clover-based topological charge of an unbiased RHMC run compared to a simulation using PT-MetaD in full QCD.}
    \label{fig:PT-MetaD_timeseries_top_qcd}
\end{figure}

\begin{figure}[H]
    \centering
    \includegraphics{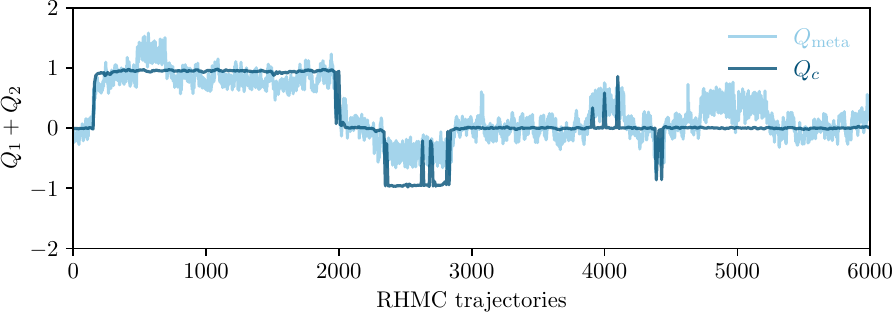}
    \caption{Time series of the summed clover-based topological charges and collective variables from both PT-MetaD streams.}
    \label{fig:PT-MetaD_timeseries_sum_qcd}
\end{figure}

With the aforementioned strategies, we were able to reduce the number of required deposited Gaussians to $\order{1000}$ updates per walker, and six walkers in total. While this may be sufficient for relatively small lattices commonly used in finite temperature simulations, there is certainly still room for improvement. 
\Cref{fig:timeseries_multiple_walkers} depicts the time series of the buildup, while \Cref{fig:MetaD_bias_qcd} shows the resulting bias potential and its SSA-modified counterpart after 5000 RHMC trajectories per walker.
\noindent
Using the modified bias potential shown in \Cref{fig:MetaD_bias_qcd}, we performed a PT-MetaD simulation under the same parameters as before to compare it with a conventional RHMC simulation. The corresponding time series of the topological charge are depicted in \Cref{fig:PT-MetaD_timeseries_top_qcd}, and the summed charges of the two PT-MetaD simulation streams can be seen in \Cref{fig:PT-MetaD_timeseries_sum_qcd}.
\noindent
The new algorithm successfully unfreezes the otherwise completely frozen system within the given number of trajectories, but the autocorrelation time of the topological charge still seems to be rather long. In particular, the measurement stream does not change the topological sector after the first 2000 trajectories, and \Cref{fig:PT-MetaD_timeseries_sum_qcd} reveals several extended periods where the system lingers between well-defined topological charges, e.g., at around 2500 or 5000 trajectories. A possible explanation for this behavior is an insufficiently thermalized bias potential, which is supported by the observation that the barrier heights of the extracted potential vary between different sectors. We have previously observed similar behavior in tests in the pure gauge case. Additionally, the small physical volume of the lattice and the associated suppression of the topological susceptibility may also partially contribute to the long autocorrelation times and explain why no configurations with $\abs{Q} > 1$ were observed.
\section{Conclusion}
The computational cost of lattice QCD simulations primarily depends on the volume, the pion mass, and the lattice spacing. While the volume scaling is close to linear, and algorithmic advancements have significantly alleviated the challenges associated with small pion masses, critical slowing down and topological freezing towards the continuum limit remain problematic. Notably, topological freezing appears to exhibit a universal behavior across different update algorithms and choices of gauge actions. Although the integrated autocorrelation times vary between different setups, the overall scaling behavior seems to be consistent and compatible with a dynamical critical exponent $z_a \approx$ \numrange{5}{6}.

Using a combination of Metadynamics and parallel tempering, it is possible to unfreeze previously frozen systems and re-enable transitions between topological sectors. The inclusion of a bias potential leads to decreased autocorrelation times, and reweighting can be avoided by using two simulation streams in a parallel tempering setup. One stream with a bias potential serves as a tunneling stream, and a second stream without a bias potential is used for measurements. As long as the physical actions and parameters on both streams are the same, the swap acceptance rates only depend on the bias potential, even when including dynamical fermions.

The algorithm is successful in four-dimensional SU(3) gauge theory, and initial tests in QCD simulations with dynamical fermions also appear promising, but the currently used bias potential seems to be suboptimal due to limited statistics. While we were already able to reduce the required buildup time of the potential, further improvements would be desirable. Future work will focus on improving and accelerating the construction of the bias potential, as well as extending simulations to more realistic parameters --- such as larger volumes, lighter quark masses, and a different action --- to assess the algorithm's speedup potential for phenomenologically relevant simulations. Additionally, we plan to explore synergies with other improved sampling methods and potential applications to theories beyond QCD.

\acknowledgments
We are grateful to Stephan Dürr, Jacob Finkenrath, and Kalman Szabo for insightful discussions. T.E. is supported by the Deutsche Forschungsgemeinschaft (DFG, grant No. HO 4177/1-1). Part of the computations were carried out on the PLEIADES cluster at the University of Wuppertal, which was supported by the Deutsche Forschungsgemeinschaft (DFG, grant No. INST 218/78-1 FUGG) and the Bundesministerium für Bildung und Forschung (BMBF).

\bibliographystyle{JHEP}
\bibliography{literature.bib}
\end{document}